# Microscopic effects in fragment mass distribution in fusion-fission reactions of light projectiles with heavy targets

T. K. Ghosh, S. Pal and P. Bhattacharya

*Saha Institute of Nuclear Physics, 1/AF Bidhan Nagar, Kolkata 700 064, India*

K. S. Golda

*Nuclear Science Centre, New Delhi-110067, India.*

(Dated: February 12, 2005)

## Abstract

Microscopic effects in fragment mass distributions in fusion-fission reactions of light projectiles (C, O and F) on deformed thorium and spherical bismuth targets in near and below Coulomb barrier energies are investigated. Precisely measured mass distribution shows a sudden anomalous increase in variances of mass distributions ($\sigma_m^2$) near Coulomb barrier energies for all three projectiles with deformed thorium target, in contrast to a smooth variation of $\sigma_m^2$ with energy for spherical bismuth target. Macroscopic effects of change in mass flow or prolonged mass equilibration time can not explain the observed variation in $\sigma_m^2$ with energy. Microscopic effects due to change in entrance channel shape compactness for projectiles hitting the polar region of prolate thorium target is postulated to reach a almost symmetric saddle without complete fusion for events of anomalous fragment widths. Quantitative estimates of mass widths mixed for the two processes explains the observed variation of $\sigma_m^2$ with excitation energy.

The evolution of a super-heavy element in fusion of two nuclei is restricted by two key factors, one is the primary fusion process to produce the super-heavy element, and the other is the radioactive decay of the fused system to survive long enough. Considerable interest and experimental studies and theoretical calculations have been done in recent years in probing the fusion and subsequent decay, mostly by binary fission, of the composite systems.

In the fusion of two heavy nuclei, the nuclei must have enough kinetic energy to overcome the repulsive electrostatic energy to be in the range of the attractive nuclear forces in a touching configuration. The fusion of the two nuclei from a touching configuration to a composite system equilibrated in all macroscopic degrees of freedom is governed by the path the system takes in a complicated multidimensional energy landscape [1] where the microscopic quantum-mechanical effects of the motion of the neutrons and protons play a prominent role and must be taken into account. The potential energy landscape in multi-dimensions depends critically on excitations and deformations of the fusing masses, the mass asymmetry, necking and the separation between two masses [2, 3]. Depending upon the initial conditions of excitations, degree of damping of radial motion, deformations and left-right mass asymmetry [$\alpha = (M_L - M_R)/(M_L + M_R)$] the system can equilibrate to a compound nucleus (reach a fusion meadow in the energy landscape). The fused system in fusion meadow may cool down with emission of a few particles and be referred as an evaporation residue (ER), or more frequently, with shape changes over a saddle ridge, slides down into a fission valley undergoing fission . However, it is also possible that the path in the energy landscape do not reach a fusion meadow, but re-separate into two fragments with altered mass asymmetry and excitation energies and deformations in a process which is tagged as a quasi-fission event.

The direct evidence of the system following a path to a fusion meadow is the observation of the ER's. In reactions in which fusion is followed by fission, the indirect evidence of the fusing system reaching a equilibrated compound nucleus is the angular distribution of the fission fragments following macroscopic statistical laws [4]. In many systems with a deformed target and/or projectile, it had been observed that at near and below Coulomb barrier energies, the angular anisotropy, defined as the ratio of the fragment yields W(0°)/W(90°), shows an anomalous increase over that predicted by the statistical models. The observed anomalous angular anisotropies are sought to be explained in the macroscopic picture. The radial motions are completely damped and excitations are assumed to reach equilibration, but K-degrees of freedom (K is the projection of the total angular momentum on the symmetry axis of the fissioning nucleus) does not reach equilibration, consequently the narrower width of the K-distributions anomalously increase the angular anisotropy of the fragments. However, the total production cross section of the ER's are not expected to be changed even if the K degrees are non-equilibrated. In another school of thought[5], in a deformed target-projectile system, at near and below Coulomb energies, fusion is assumed to be dependent on the relative orientation of the symmetry axes of the target and projectile. For the polar region of impact, the microscopic effects of the interactions of nucleons changes the path in the energy landscape and the system reaches a fission valley without crossing the fusion meadow, in the so called orientation dependent quasi-fission reaction. In this process, the production of the ER's are necessarily hindered.

A critical test for the two possibilities are a direct measurement of the ER's. Hinde et al reported hindrance in fusion of $^{220}$Th with different entrance channel target and projectiles of decreasing mass asymmetry [6]. Similar observations were reported in the fusion of $^{216}$Ra [7]. With matching of the excitation energies and angular momentum, the ER productions are claimed to be hindered with increasing deformation (spherical $^{204}$Pb, oblate $^{197}$Au and prolate $^{186}$W) of the target and also with macroscopic effect of the direction of mass flow in the fusion process. A compact spherical system is more probable from a mass asymmetric initial state ( $^{12}$C + $^{204}$Pb, $^{16}$O + $^{204}$Pb) and more hindered for more mass symmetric pairs ($^{40}$Ar + $^{180}$Hf, $^{48}$Ca + $^{172}$Yb, $^{82}$Se + $^{138}$Ba, $^{124}$Sn + $^{96}$Zr) purely from the closeness of the initial and final states. But microscopic effects (more binding energies of Pb and Se) [1] should also be considered in considering the ER productions. Recently, completely conflicting experimental evidence on the ER production in the deformed target of $^{238}$U with $^{16}$O has been reported [8]. It has been shown that even for projectiles hitting the polar region of the uranium nuclei, the fusion of the oxygen nuclei are unhindered when compared with those cases where the oxygen hits the equatorial region of deformed uranium nuclei, although angular anisotropy of fragments are reported to be anomalously large compared to macroscopic predictions [9]. So it is clear that more experimental probes, other than the angular anisotropy or the production of ER's are needed to understand the path of fusion of two nuclei in the energy landscape leading to different processes.

The departures from the fusion-fission paths governed by macroscopic nuclear effects are possibly due to microscopic effects changing the initial conditions for the fusion process.

Depending upon initial conditions, it may be conjectured that the fusion-fission process may follow different paths to reach the fission valley due to microscopic effects and may be differing in exit channel mass asymmetries. So precise measurements of exit channel mass distributions and changes in the mass distributions with macroscopic parameters, e.g., excitation energy of the fused system, may be a good probe to study the dynamics of the fusion-fission process.

In the experimental studies, we have determined precisely the fission fragment mass distributions in reactions of spherical or very slightly deformed light projectiles of $^{19}$F, $^{16}$O and $^{12}$C on spherical $^{209}$Bi and deformed $^{232}$Th targets in near and below Coulomb barrier energies. In the systems with the spherical bismuth target, the entrant system is compact for any orientation and the mass flows are from target to projectile in all target projectile systems [10]. However, the entrance channel compactness in shape changes quite appreciably for the impact point of the projectile changing from equatorial to polar regions of the prolate thorium nuclei. Also the macroscopic effects of mass flow for carbon is opposite to that of oxygen and fluorine nuclei reacting with thorium target. In all the cases, the macroscopic effects only predict a smooth variation of the width of the fragment mass distributions with the excitation energies or the temperature of the equilibrated fused system [11]. So any departure of the smooth variation of the width of the mass distributions are looked for in the experimental studies.

The experiments were performed with pulsed heavy ions of $^{12}$C, $^{16}$O and $^{19}$F from the 15UD Pelletron at Nuclear Science Centre (NSC), New Delhi, India. The pulse width was about 0.8-1.5 ns with a pulse separation of 250 ns. The targets were either self-supporting $^{232}$Th of thickness 1.8 mg/cm$^2$ or a 500 $\mu$g/cm$^2$ thick self-supported $^{209}$Bi. Complimentary fission fragments were detected with two large area (24 cm $\times$ 10 cm) X-Y position sensitive multi-wire proportional counters (MWPCs) [12]. The detectors were placed at 52.6 cm and 33.2 cm from the target. The fission fragments were separated from elastic and quasi-elastic channels using time of flight of particles and the energy loss signal in the detectors. Folding angle technique [13] was used to differentiate between fusion-fission (FF) and transfer fission (TF) events. The masses of the fission fragments were determined event by event from precise measurements of flight paths and flight time differences of complimentary fission fragments. The estimated mass resolution for fission fragment was about 3 a.m.u. The details of experimental arrangements and data analysis and elimination of systematic errors were reported in ref [12, 14].

The measured mass distributions in earlier reported cases of $^{19}$F, $^{16}$O + $^{232}$Th and $^{16}$O +$^{209}$Bi [14, 15] and the presently reported in case of $^{12}$C + $^{232}$Th and $^{19}$F + $^{209}$Bi at all energies are well fitted with single Gaussian distributions around the symmetric mass split for the target plus projectile systems. The variation of the square of the variance of the fission fragment mass distribution ($\sigma_m^2$) are shown in FIG 1 for $^{19}$F and $^{16}$O projectiles on the spherical bismuth nuclei. It has been observed that the mass variance ($\sigma_m^2$) shows a smooth variation with the excitation energy of the fused system across the Coulomb barrier. This is in qualitative agreement with the predictions of statistical theories. It is also noted that no significant departures are reported in the fragment angular anisotropy measurements for the spherical target and projectile systems [15–17].

The variances of mass distribution ($\sigma_m^2$) for reactions of different projectiles for the present as well as earlier reports [14, 15] on the deformed Thorium target are shown in FIG 2, FIG 3 and FIG 4 for $^{19}$F, $^{16}$O, $^{12}$C projectiles respectively. In all three cases, as the excitation energy is decreased, the $\sigma_m^2$ values decrease monotonically, but shows a sudden upward trend approximately around the Coulomb barrier energies. This is once again followed by a smooth decreasing trend as energy is further decreased. The sudden increase in ($\sigma_m^2$) values is most prominent ($\sim 50\%$) in case of $^{19}$F + $^{232}$Th and decreases to ($\sim 15\%$) in $^{16}$O + $^{232}$Th and to ($\sim 10\%$) the in $^{12}$C + $^{232}$Th system. It has been simulated and experimentally verified that rise in ($\sigma_m^2$) values can not be explained by any systematic error, e.g., loss of energy of fragments in target or mismatch of timing in two T.O.F. arms. It is interesting to note that anomalous increase in the angular anisotropy of fission fragments were observed in almost at the same beam energies at which anomalous increase in fragment angular anisotropies were observed in all systems [18].

Observance of a sudden rise in ($\sigma_m^2$) values as the excitation energy is lowered may signify a mixture of two fission modes, one following the normal statistical predictions of fusion-fission path along zero left-right mass asymmetry ($\alpha$), and another following a different path in the energy landscape with zero or small mass asymmetry. The mixture of the two modes may give rise to wider mass distributions. Similar to the postulation of the orientation dependent quasi-fission [5], we postulate that for fusion-fission paths corresponding to the projectile orientations up to a critical angle ($\theta_c$) of impact on the polar region of prolate thorium, the width and energy slope of the symmetric mass distributions are different (shown by dot-dashed curves in FIG2,FIG3 and FIG4) compared to those for the normal statistical fusion-fission paths (dotted curves). The mass widths weighted by the fission cross sections ( which are assumed to be very close to fusion cross section as the composite systems are of high fissility) from earlier measurements [18] are mixed for the two fusion-fission modes and shown by different coloured continuous curves for different critical polar angles separating the two fission modes, for all three systems. The calculated ($\sigma_m^2$) values quantitatively explain the observed increase in the widths of the mass distributions. It is interesting to note that the fusion-fission process is clearly dominated by the normal process at above Coulomb barriers and the "anomalous" fusion-fission process is dominant at lower energies. However, experimental evidence suggests that the variations of mass distributions with excitation energies are similar for the both processes, probably dominated by macroscopic forces, but differing quantitatively due to microscopic effects.

We have clearly established with the present string of measurements that widths of the mass distributions is a sensible tool to observe departure from the normal fusion-fission path in the fusion of heavy nuclei. The exact mechanisms for the departure from normal fusion-fission paths are not known accurately. However, macroscopic effects such as the direction of mass flow or the mass relaxation time being too prolonged may not be the cause. It has been established earlier from the experimental barrier distributions, the reaction cross sections in $^{19}$F, $^{16}$O, $^{12}$C + $^{232}$Th in near and below Coulomb barrier energies are mostly for impact of the projectiles on the polar regions of the thorium nuclei. Following the quantum mechanical effects favouring similar shapes in entrance and exit channels [1], we modify the simple postulation of the microscopic effects of the relative orientation of the projectile to the nuclear symmetry axes of the deformed target [5]. We assume that for the non-compact entrance channel shape, the impact of the projectile in the polar region of $^{232}$Th target drives the system to an almost mass symmetric saddle shape, rather than a compact equilibrated fused system. The observed fragment mass widths can be quantitatively explained under such assumptions. The above postulation is supported by the observation that for the spherical target $^{209}$Bi, where entrance channel compactness of shape is same for all relative target-projectile orientations, only normal fusion-fission paths, as characterized by the smooth variation of fragment mass widths with excitation energy, are observed. It is also worthwhile to note that effect of the anomalous mass widths increases with left-right mass symmetry in the entrance channel in case of $^{19}$F, $^{16}$O, $^{12}$C + $^{232}$Th system in consonance with our postulation.

The present observations highlight the need for very detailed microscopic calculations of the dynamics of the nucleons in the dissipative system. Extensive calculations of the multidimensional potential energy surface has successfully explained spontaneous and low energy fission phenomena [2, 3]. Calculated paths through the minimum energy valleys and over ridges in the potential surface showed that apart from the deformations and necking of the two nascent fragments, the left-right mass asymmetry also plays a crucial role. All the heavier than actinide nuclei show a mass symmetric ($\alpha = 0$) and mass asymmetric ($\alpha \neq 0$) saddle shapes with a ridge separating the two down the scission path. The relative heights of the two saddles and the separating ridge governed the symmetric, asymmetric or a mixture of the two fission paths in specific cases. Recent extensions [19, 20] of the five dimensional energy landscapes for the fusion in a dynamical calculation through a dissipating system have been carried out in the case of fusion of heavy nuclei in above Coulomb barriers . It has been observed that the saddle shapes are symmetric for the fusion-fission (FF) path, but depending upon the initial conditions of mass asymmetry and dissipation of the radial motion, most of the paths may deviate through a mass asymmetric saddle shape before fusion to re-separate in a QF reaction mode. In a very similar situations, the observed microscopic effects in the mass distributions may be due to subtle changes in the initial conditions of the motion of the nucleons in a dissipative system brought in by the difference in compactness of the entrance channel shape and the possible changes in fusion and heights and separation of mass symmetric fission valleys and ridges. This calls for more detailed calculations of the microscopic effects in fusion processes.

Authors would like to thank the staff at NSC Pelletron for providing excellent beam and other logistical support and help during the experiment. Help of Drs. A. Saxena, D. C. Biswas, S.Chattopadhyay, Mr. P. K. Sahu during the experiments and discussions with Drs R.K.Bhowmick and S.K.Datta are gratefully acknowledged.

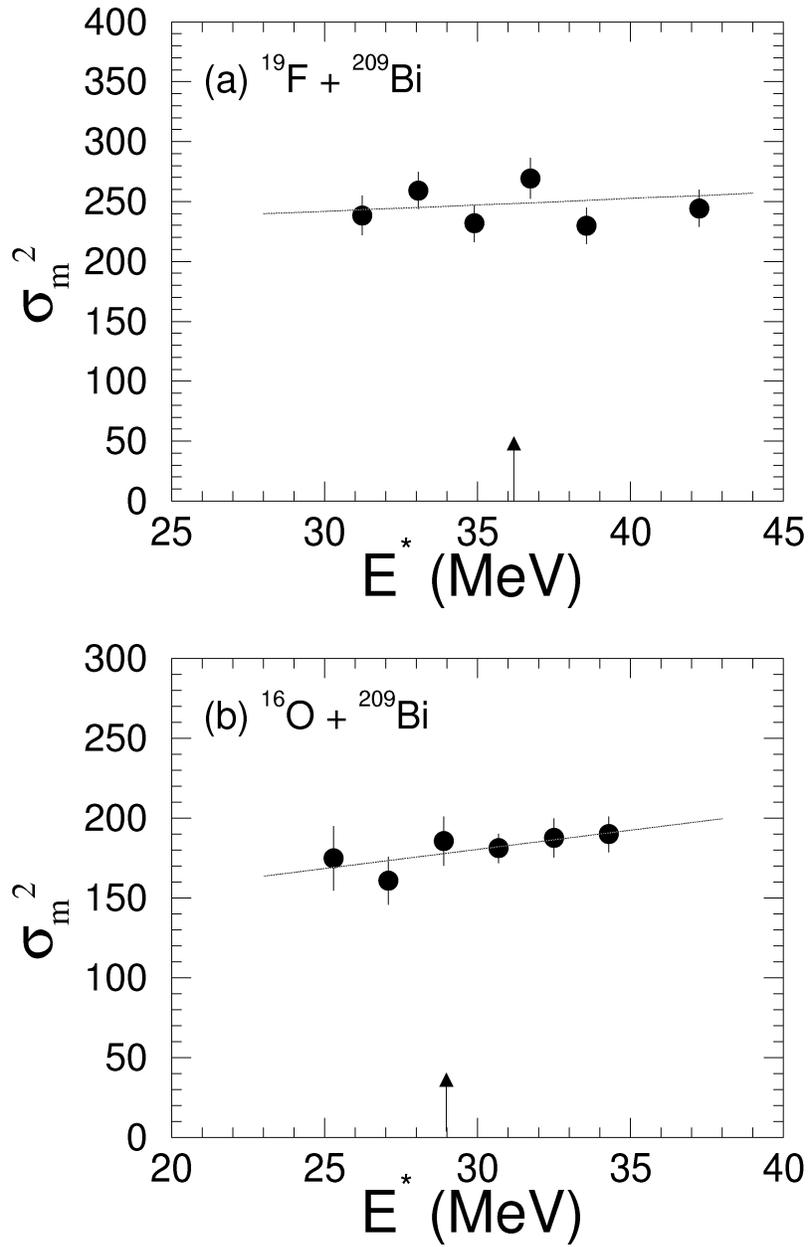

FIG. 1: Mass variance ($\sigma_m^2$) as a function of excitation energy (E⋆) for spherical bismuth target. The arrow points to excitation energy corresponding to Coulomb barrier. The solid lines (red) show smooth variation of $\sigma_m^2$ with E⋆

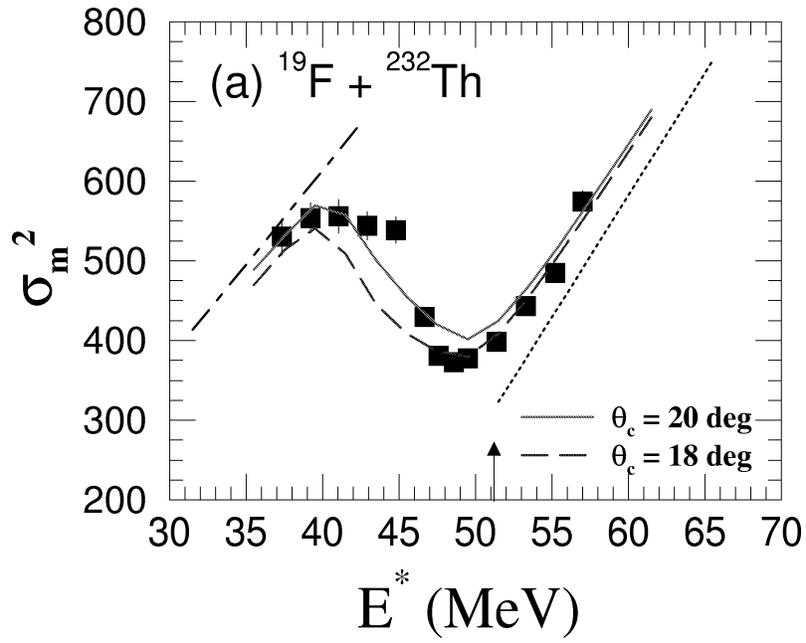

FIG. 2: Variation of $\sigma_m^2$ with excitation energy for the system $^{19}$F + $^{232}$Th. The dotted and dot-dashed curves are postulated variation for normal and "anomalous" fission modes. Calculated $\sigma_m^2$ for two critical angles ($\theta_c$) are indicated

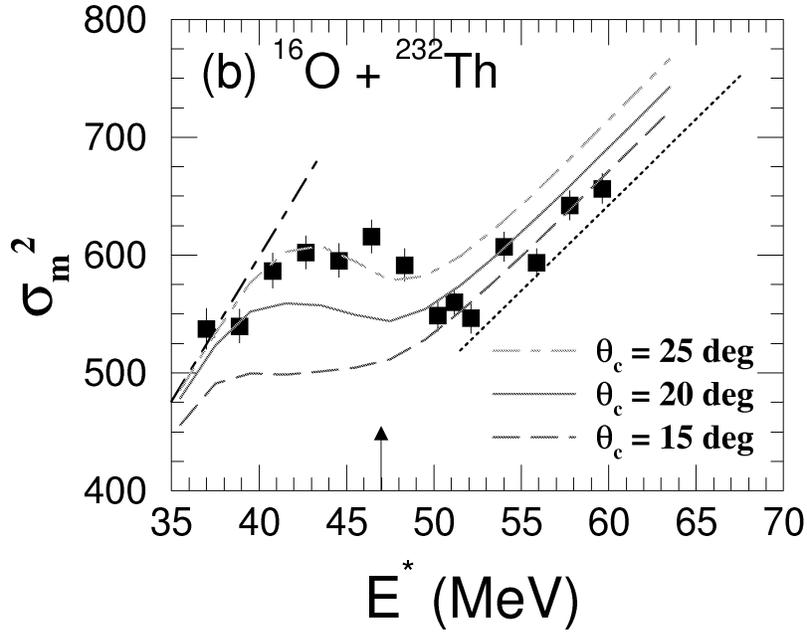

FIG. 3: Variation of $\sigma_m^2$ with excitation energy for the system $^{16}$O+$^{232}$Th

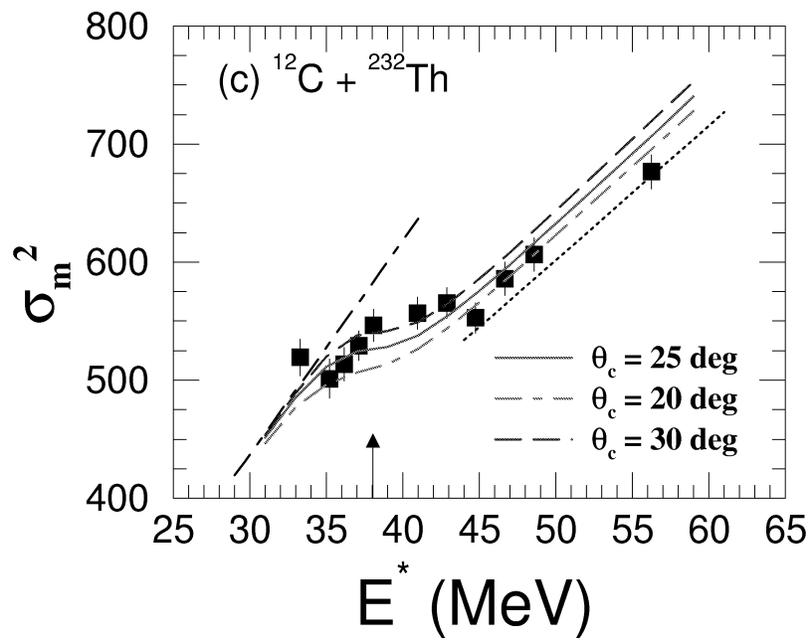

FIG. 4: Variation of $\sigma_m^2$ with excitation energy for the system $^{12}$C+$^{232}$Th